\documentclass[12pt, english]{iopart}
\usepackage[nofoot]{geometry}
\usepackage{graphicx}
\usepackage[center]{subfigure}
\usepackage{iopams}

\begin{document}

\title[Ohm's law in reduced MHD]{On Ohm's law in reduced plasma fluid models}
\author{B.D.Dudson$^1$, S.L.Newton$^2$, J.T.Omotani$^2$ and J.Birch$^{2,3}$}

\address{$^1$ York Plasma Institute, Department of Physics, University of York, Heslington, York YO10 5DQ, UK} 
\address{$^2$ CCFE, Culham Science Centre, Abingdon, Oxon OX14 3DB, UK}
\address{$^3$ University of Exeter, Stocker Rd, Exeter EX4 4PY}

\ead{benjamin.dudson@york.ac.uk}

\begin{abstract}
Drift-reduced MHD models are widely used to study magnetised plasma phenomena, in particular
for magnetically confined fusion applications, as well as in solar and astrophysical research.
This letter discusses the choice of Ohm's law in these models, the resulting dispersion relations
for the dynamics parallel to the magnetic field, and the implications for numerical simulations.
We find that if electron pressure is included in Ohm's law, then both electromagnetic and
finite electron mass effects must also be included in order to obtain physical dispersion relations.
A simple modification to the plasma vorticity is also found which improves handling of low density
regions, of particular relevance to the simulation of the boundary region of magnetised plasmas.
\end{abstract}


\section{Introduction}

Drift-reduced fluid models are widely used for the study of low-frequency (relative to the ion cyclotron frequency) plasma phenomena, in relatively collisional regimes where fluid models are appropriate~\cite{braginskii-1}, such as the edge region of present-day tokamaks. These models have also been applied to the solar corona~\cite{buchlin2007,ballegooijen2011} and interplanetary turbulence~\cite{schekochihin2007}.
A wide range of models have been derived in the literature, see for example~\cite{rosenbluth1976,strauss1976,strauss1982,mikhailovskii1984,drake1984,hazeltine1985,hsu1986,hazeltine1987,pfirsch1996,scott1997,kruger1998,simakov-2003,brizard2005,ramos2005,reiser2012}. The basic assumption they share is that
at low frequencies the cyclotron motion can be averaged over, and the plasma fluid motion perpendicular to the magnetic field described by drifts due to magnetic and electric field
inhomogeneties. The electrostatic potential is often determined by enforcing quasineutrality, whilst in electromagnetic models the magnetic field
perturbation is determined through Amp\'ere and Ohm's law. 

Most models are constructed to conserve an energy, either by careful selection of terms in an ordering expansion (e.g. \cite{hsu1986,simakov-2003}), or by deriving the equations from a 
Lagrangian (see e.g.\cite{pfirsch1996,brizard2005}). Whilst energy conservation is important for both linear and nonlinear properties~\cite{pfirsch1996,scott1997}, and usually improves the numerical stability
of a model implementation, it is not the only consideration. The dispersion relation of the model,
and the characteristics of the waves it supports, are less often detailed. Physically the group speed of these waves determines how fast information propagates in the system, which must not exceed the speed of light in vacuum. 

In numerical implementations, the fastest waves set the CFL limit on explicit time steps, and 
for implicit time stepping methods~\cite{rognlien-2002,schnack2005,jardin2012} contributes to the difficulty of inverting the system Jacobian. 
Implementations of 3D drift-reduced fluid models for plasma turbulence applications include BOUT~\cite{xu-2008,umansky-2008-bout}, BOUT++ \cite{Dudson2009}, GBS~\cite{ricci2012}, TOKAM-3D~\cite{tamain2010}, and STORM~\cite{easy2014}. 
Each of these codes use different numerical methods, coordinate systems, and have variations in the form of equations solved, but they all share
common features of drift-reduced systems: a vorticity equation, coupled to a parallel Ohm's law and a density continuity equation. Here we examine numerical issues arising from the
the linear behavior of this fundamental set of equations. 

This letter aims to clarify discussion of the physical completeness and numerical stability of reduced plasma fluid models, and in particular the effect of the choice of Ohm's law. 
Several studies employing reduced fluid models have retained both electromagnetic and electron inertia effects (e.g. \cite{scott1997,ricci2012,cohen2013,lee2015}) and there have been discussions of the resulting dispersion relation \cite{hazeltine1985,scott1997,schnack2005}, but many studies do not include these terms, and the impact on the dispersion relation of the various approximations has not to our knowledge been clearly presented in one place. 
Therefore in section~\ref{sec:results} we introduce the drift-reduced equations, then examine the behaviour of the electrostatic approximation in section~\ref{sec:estatic} and move on to consider electromagnetic
models in section~\ref{sec:emag}. Section~\ref{sec:benchmark} illustrates the practical importance of the linear modes for timestep in example nonlinear simulations. Conclusions are given in section~\ref{sec:conclusion}.

\section{Drift-reduced MHD model dispersion relations}
\label{sec:results}

In this letter we consider a series of simplified drift-reduced plasma fluid models, which evolve
the (electron) density $n$, vorticity $U$, and take a form for Ohm's law parallel to the magnetic field.
A full derivation of the drift-reduced equations can be found elsewhere~\cite{simakov-2003} and is
beyond the scope of this paper.
We focus for simplicity on perturbations of a uniform stationary pure plasma in a uniform magnetic field, in the cold ion limit, so we do not need to include finite Larmor radius corrections to the drift motion of either species or ion viscosity.
We give a brief outline of the derivation,
using SI units throughout except for electron temperature $T_e$ in eV, which we follow by considering step-by-step the form of the Ohm's law.

The electron density continuity equation is used to avoid the need to evaluate high order ion polarisation velocity terms:
\begin{equation}
\frac{\partial n}{\partial t} = -\nabla\cdot\left(n\mathbf{v}_e\right),
\end{equation}
where ${\bf v}_e$ is the electron fluid velocity.
The quasineutrality assumption is enforced through current continuity
\begin{equation}
\nabla\cdot\mathbf{J} = 0,
\label{eq:div_j}
\end{equation}
and the current parallel to the magnetic field $J_{||} = \mathbf{b}\cdot\mathbf{J}$ is derived from the electron momentum equation~\cite{braginskii-1}
\begin{equation}
m_en\frac{d\mathbf{v}_e}{dt} + \nabla p_e + \nabla\cdot\Pi_e + en\left(\mathbf{E} + \mathbf{v}_e\times\mathbf{B}\right) = \mathbf{F},
\label{eq:emomentum}
\end{equation}
where $\mathbf{F}$ is the friction between electrons and ions, which will appear in the equations below as resistivity.
In the drift approximation it is assumed that charged particle motion is due to parallel flow along magnetic field-lines, and drifts
in the perpendicular direction. For simplicity we here neglect $E\times B$ drift (since it does not lead to charge separation) and as we
assume constant background magnetic field, the diamagnetic drift is also neglected.
The relevant velocity of electrons and ions would therefore be given here by the sum of parallel flow and polarisation drift: $\mathbf{v}_{i,e} = \mathbf{b}v_{||i,e} + \mathbf{v}^{pol}_{i,e}$.
Finally, following standard orderings, the electron polarisation drift and viscosity $\Pi_e$ are neglected, as they are small in the electron mass.
The ion polarisation drift is approximately given by
\begin{equation}
\mathbf{v}^{pol}_i \simeq \frac{m_i}{eB^2}\frac{d\mathbf{E}_\perp}{dt},
\end{equation}
which arises from the time-derivative of the $E\times B$ flow velocity.

Using the Coulomb gauge, $\mathbf{E} = -\nabla\phi - \partial_t\mathbf{A}$, but following the standard low-$\beta$ orderings 
the electromagnetic contribution to $\mathbf{E}_\perp$ is neglected, so $\mathbf{E}_\perp \simeq -\nabla_\perp\phi$.
Since we are interested in the linear response only, with a stationary background, the nonlinear advection term can be dropped
so that
\begin{equation}
\frac{d\mathbf{E}_\perp}{dt} \simeq -\frac{\partial}{\partial t}\nabla_\perp\phi.
\end{equation}
Similarly we can neglect here the variation of density in the ion polarisation, sometimes called the Boussinesq
approximation (see~\cite{angus2014} and references therein) when the background is not formally uniform, so the divergence of the ion polarisation current can be written as
\begin{equation}
\nabla\cdot\mathbf{J}^{pol}_i = \nabla\cdot\left(en\mathbf{v}^{pol}_i\right) \simeq -\frac{m_i n_0}{B^2}\frac{\partial}{\partial t}\nabla_\perp^2\phi,
\end{equation}
where $n_0$ is a constant background density.

Thus we have the minimal set of reduced MHD equations which we will study in this letter:
\begin{eqnarray}
\frac{\partial n}{\partial t} &=& \nabla\cdot\left(\mathbf{b}\frac{J_{||}}{e}\right), \label{eq:density} \\
\frac{m_i n_0}{B^2}\frac{\partial }{\partial t}\nabla_\perp^2\phi &=& \nabla\cdot\left(\mathbf{b}J_{||}\right), \label{eq:vorticity} \\
\frac{m_e}{e}\frac{\partial v_{||e}}{\partial t} &=& -\frac{1}{en}\partial_{||}p_e +\partial_{||}\phi + \frac{\partial A_{||}}{\partial t} + \eta J_{||}, \qquad  \label{eq:ohms_law} \\
\frac{m_i}{e}\frac{\partial v_{||i}}{\partial t} &=& -\partial_{||}\phi - \frac{\partial A_{||}}{\partial t} - \eta J_{||}, \qquad  \label{eq:ion_moment} \\
&& J_{||} = -\frac{1}{\mu_0}\nabla^2A_{||},
\end{eqnarray}
which are essentially the same equations as used in \cite{angus2014,lee2015,walkden2015}. Remember in the isothermal electron limit the thermal force does not appear in the expression for the parallel friction ${\bf b}\cdot {\bf F} = en_0\eta J_\parallel$, where the parallel Spitzer resistivity for a pure deuterium plasma is $\eta = 0.51m_e/n_0e^2 \tau_{ei}$ and $\tau_{ei} = 12 \pi^{3/2}m_e^{1/2}T_e^{3/2}\epsilon_0^2/2^{1/2}n_0 e^{5/2} \ln \Lambda$ is the electron-ion collision time~\cite{perbook}. Note that~\cite{pfirsch1996}, for example, gives a more complete expression for the parallel electron velocity, which reduces to equation~\ref{eq:ohms_law} in the linear limit we will consider here.

The inclusion of electron inertia or pressure in Ohm's law (equation~\ref{eq:ohms_law}) significantly change
the character of the waves in this system, as compared to ideal or resistive MHD, introducing
the inertial (IAW) or kinetic Alfv\'en waves (KAW)~\cite{uberoibook}, whose wave speeds along the magnetic field have a dependence on wave-number perpendicular to the magnetic field, $k_\perp$.
In the following sections we examine the effect of approximations made to equation~\ref{eq:ohms_law} on the dispersion relations, and do not note explicitly the various $\omega=0$ solutions which arise.
Here we are concerned with ensuring that the resulting wave speeds do not diverge or exceed the speed of light in vacuum $c$
for arbitrary $k_\perp$.


\subsection{Electrostatic model}
\label{sec:estatic}

In the case of isothermal electrons with temperature $T_e$
\begin{equation}
\frac{1}{en}\partial_{||}P_e \rightarrow \frac{T_e}{n_0}\partial_{||}n.
\end{equation}
Neglecting both the electromagnetic part of the electric field and the electron mass, Ohm's law (\ref{eq:ohms_law}) then reduces to
\begin{equation}
\eta J_{||} = -\partial_{||}\phi + \frac{T_e}{n_o}\partial_{||}n,
\end{equation}
so that in the limit of $\eta \rightarrow 0$ the well known Boltzmann relationship between potential $\phi$ and density $n$ is recovered. 
If we also neglect the ion parallel flow, and so exclude the parallel ion sound wave, the parallel current can be written as $J_{||} = -env_{||e}$.
Linearising the set of equations~\ref{eq:density}-\ref{eq:ohms_law}, with
\begin{equation}
\partial_{||}\rightarrow ik_{||}, \qquad \frac{\partial}{\partial t} \rightarrow -i\omega, \qquad \nabla_\perp^2 \rightarrow -k_\perp^2,
\end{equation}
the dispersion relation with finite resistivity is then given by
\begin{equation}
-i\omega = -k_{||}^2\frac{T_e}{\eta e n_0}\left(\frac{1}{k_\perp^2\rho_s^2} + 1\right).
\label{eq:static_zem}
\end{equation}
Here $\rho_s = c_s/\Omega_i = \sqrt{T_em_i/eB^2}$ is the hybrid ion or ion sound gyroradius, where $c_s = \sqrt{eT_e/m_i}$ is the ion sound speed and $\Omega_i$ is the ion cyclotron frequency. Typical magnetised plasma edge turbulence, for example in tokamaks and 
linear devices, have scale lengths perpendicular to the magnetic field such that $k_\perp\rho_s \sim 0.1 - 1$.

Equation~\ref{eq:static_zem} represents a diffusion equation along the magnetic field for the evolving quantities ($n$, $\phi$, or $J_{||}$) with diffusion coefficient $D$:
\begin{equation}
\qquad D = \frac{T_e}{\eta e n_0}\left(\frac{1}{k_\perp^2\rho_s^2} + 1\right).
\end{equation}
 Smaller $k_\perp$ modes (longer perpendicular wavelength) diffuse
along the magnetic field faster than high $k_\perp$ modes.
With the Spitzer resistivity $\eta \simeq 10^{-4} \ln \Lambda T_e^{-3/2}~ \Omega$m for a pure deuterium plasma:
\[
D \simeq 1.3\times 10^6 \left(\frac{T_e}{10\mathrm{eV}}\right)^{5/2}\left(\frac{10^{18}\mathrm{m}^{-3}}{n_0}\right)\left(\frac{1}{k_\perp^2\rho_s^2} + 1\right) ~ \textnormal{m}^2/\textnormal{s}.
\]
This fast diffusion can place limits on simulation time steps, 
becoming more restrictive as the electron temperature increases or the system size increases (smallest $k_\perp$ decreases). Note that there is no limit to the speed at which information propagates in this model: $k_\perp = 0$ modes communicate instantaneously along magnetic field lines.

If we keep finite electron mass in Ohm's law, but still drop the electromagnetic term then equation~\ref{eq:ohms_law} becomes
\begin{equation}
\eta J_{||} = -\partial_{||}\phi + \frac{1}{en}\partial_{||}P_e + \frac{m_e}{e}\frac{\partial v_{||e}}{\partial t},
\end{equation}
giving the dispersion relation
\begin{equation}
\omega^2 + i\omega\eta\frac{v_{te}^2}{\mu_0V_A^2\rho_s^2} = k_{||}^2v_{te}^2\left(1 + \frac{1}{k_\perp^2\rho_s^2}\right),
\label{eq:static_fem}
\end{equation}
where $v_{te} = \sqrt{eT_e/m_e}$ is the electron thermal speed. The second term on the left here accounts for perpendicular magnetic diffusion due to the parallel resistivity, which results from the electron-ion collisions which have a characteristic timescale $0.51/\tau_{ei} = \eta n_0 e^2 / m_e = \eta \omega_{pe}^2/\mu_0 c^2 =  \eta v_{te}^2/\mu_0V_A^2\rho_s^2$. The plasma skin depth is $c/\omega_{pe}$, with $c$ the speed of light in vacuum, $\omega_{pe} = \sqrt{n_0e^2/\epsilon_0 m_e}$ the plasma frequency, while the Alfv\'{e}n speed $V_A=B/\sqrt{\mu_0m_in_0}$ is introduced in the last form by noting the characteristic perpendicular lengthscale of the system is the sound gyroradius. Rather than a diffusion equation, as~\ref{eq:static_zem}, the system dispersion relation is now a wave equation.  
Writing the solution as $\omega = \omega_0 + i \gamma$, where $\omega_0$ and $\gamma$ are real,
\begin{equation}
\hspace{-2cm} \omega_0 = \frac{k_\parallel}{k_\perp}\frac{v_{te}}{\rho_s}\sqrt{1+ k_\perp^2 \rho_s^2 - \frac{\eta^2}{4}\frac{k_\perp^2}{k_\parallel^2}\frac{v_{te}^2}{\mu_0^2V_A^4\rho_s^2}}, \qquad \gamma = - \frac{\eta}{2}\frac{v_{te}^2}{\mu_0V_A^2 \rho_s^2} = - \frac{0.51}{2\tau_{ei}}.
\label{eq:static_femsoln}
\end{equation}
As we are considering the fluid limit, the collision frequency must be faster than the other timescales in the system, so these waves must damp rapidly. 
The same problem is encountered however: 
the wave speed diverges as $k_\perp \rightarrow 0$.

Neglecting resistivity, in the cold plasma limit $T_e = 0$, the dispersion relation~\ref{eq:static_fem}, reduces to
\begin{equation}
\omega^2 = \Omega_i^2\frac{k_\parallel^2}{k_\perp^2}\frac{m_i}{m_e}.
\label{eq:static_esgk}
\end{equation}
This is the electrostatic wave which has been noted in the context of gyrokinetic simulations to limit the timestep, see for example the early discussion by Lee~\cite{lee1987} or a recent case by McMillan~\cite{mcmillan2020}.

These fast waves present a problem for global plasma simulations, where the largest perpendicular scale is much greater than $\rho_s$: As the system size increases, the fastest wave speed will rapidly increase, and for (semi-)implicit methods the problem will become increasingly poorly conditioned.

\subsubsection{Ion parallel momentum}

If the ion parallel momentum is also evolved, in the electrostatic approximation, equation~\ref{eq:ion_moment} gives
\begin{equation}
  \frac{m_i}{e}\frac{\partial v_{||i}}{\partial t} = -\partial_{||}\phi - \eta J_{||}.
\end{equation}
and the dispersion relation becomes
\begin{equation}
\omega^2+i \omega \eta \frac{v_{te}^2}{\mu_0V_A^2\rho_{sm}^2}\left(1-\frac{k_\parallel c_{sm}^2}{\omega^2}\right)=\frac{k_\parallel^2v_{te}^2}{k_\perp^2\rho_{sm}^2}\left(1+k_\perp^2\rho_{sm}^2 - \frac{k_\parallel^2c_{sm}^2}{\omega^2}\right),
\label{eq:static_femionmom}
\end{equation}
where for convenience we have defined the electron mass corrections $c_{sm}^2 = c_s^2/(1+m_e/m_i)$ and $\rho_{sm} = c_{sm}/\Omega_i$.
Neglecting the electron mass, in the limit of zero resistivity, this reduces to simply the bracket on the right equal to zero
\[
\omega^2 = k_{||}^2c_s^2\frac{1}{\left(1 +  k_\perp^2\rho_s^2\right)},
\]
and we see that we have introduced the ion acoustic wave $\omega^2 = k_{||}^2c_s^2$ to the system, including finite sound radius corrections when $k_\perp \neq 0$. This wave can be destabilised in the presence of background plasma gradients to give the slab branch of drift waves.

Neglecting only the electron mass in~\ref{eq:static_femionmom}, we see the ion sound radius couples the diffusive mode and the ion acoustic wave at finite $k_\perp$. When $\omega^2 \gg k_\parallel^2 c_s^2$ the parallel ion momentum equation can be neglected, and we recover $-i \omega \eta = - \mu_0k_\parallel^2V_A^2\rho_s^2(1+k_\perp^2 \rho_s^2)/k_\perp^2 \rho_s^2$, that is the diffusive mode, equation~\ref{eq:static_zem}. By comparing the diffusive timescale to the inverse ion acoustic frequency, we see that this limit corresponds as expected to the case of $m_i\rightarrow\infty$, or $k_\perp \rightarrow 0$ at finite $k_\parallel$.

The parallel ion momentum evolution modifies the electrostatic wave described by~\ref{eq:static_esgk} simply by multiplying the right hand side of that relation by $(1+m_e/m_i)$. The modified frequencies of the resistive wave solution~\ref{eq:static_femsoln} are
\[
\hspace{-2cm}\omega_0 = \frac{k_\parallel}{k_\perp}\frac{v_{te}}{\rho_s}\sqrt{1+ k_\perp^2 \rho_s^2 + \frac{m_e}{m_i} - \frac{\eta^2}{4}\frac{k_\perp^2}{k_\parallel^2}\frac{v_{te}^2}{\mu_0^2V_A^4\rho_s^2}\left(1 + \frac{m_e}{m_i}\right)^2}, \hspace{0.5cm} \gamma = - \frac{0.51}{2\tau_{ei}}\left(1 + \frac{m_e}{m_i}\right).
\]
Thus the parallel ion momentum evolution has not removed the difficulty as $k_\perp \rightarrow 0$.






Several other physical effects could limit the rate at which these electrostatic waves propagate. For example, we have made an isothermal approximation in relating electron pressure $p_e$ to density $n$, which would break down for sufficiently fast phenomena - but we should remain in the collisional limit. We have considered the cold ion limit for simplicity, but even in the edge plasma the ion temperature will typically not be smaller than $T_e$ by the mass ratio~\cite{elmore2012,allan2016}, so finite ion Larmor radius effects should introduce further dispersion. Here however we focus on Ohm's law, where a partial solution to the problem of unphysically fast waves is known to result from retaining electromagnetic effects~\cite{lee1987}. This introduces the Alfv\'en wave and is discussed in the next section. 

\subsection{Electromagnetic model}
\label{sec:emag}

Now we examine the dispersion relation of the system retaining the electromagnetic contribution to the parallel electric field.
Again we begin by neglecting the finite electron mass, but keeping here the parallel ion momentum evolution, which gives the dispersion relation
\begin{equation}
\hspace{-1.5cm} \omega^2 + i\omega \eta\frac{k^2}{\mu_0}\left(1 - \frac{k_\parallel^2c_s^2}{\omega^2}\right)= k_{||}^2V_A^2\left[\frac{k^2}{k_\perp^2} + k^2\rho_s^2+\frac{k_\parallel^2c_s^2}{\omega^2}\left(\frac{\omega^2}{k_\parallel^2 V_A^2} -\frac{k^2}{k_\perp^2}\right)\right].
\label{eq:emag_zem}
\end{equation}
Note that this relation does not diverge as $k_\perp \rightarrow 0$, unlike the
electrostatic relation in equation~\ref{eq:static_zem}. Including the electromagnetic term can therefore improve numerical stability.
However, the equation now has a problem at high $k_\perp$: the parallel wave speed increases with $k_\perp$. Neglecting the resistivity and taking low $\beta = c_s^2/V_A^2$ so the ion acoustic wave can be neglected, we see that this is due to the kinetic Alfv\'{e}n wave with dispersion relation
\begin{equation}
    \omega^2 = k_{||}^2V_A^2\left(1 + k_\perp^2\rho_s^2\right);
    \label{eq:kaw}
\end{equation}
the origin of the $k_\perp^2\rho_s^2$ term is the $\partial_{||}P_e$ term in Ohm's law.

Typical blob or turbulence simulations need to resolve smallest scales of around
$\delta_x \sim \rho_s$, so the highest $k_\perp$ in the simulation has $k_\perp = 2\pi/\delta_x \sim 10/\rho_s$. Adding this term therefore
introduces a wave into the system around $10$ times faster than the Alfv\'en speed. 
For deuterium plasmas in a 1T magnetic field, this wave will exceed the speed of light once the density falls below $n_0 \sim 2.6\times 10^{17}$m$^{-3}$.
This can quite easily occur in plasma edge simulations whose domain
often includes a near-vacuum region such as the far scrape-off layer (SOL) in tokamak simulations. 


Going on to include finite electron mass,
the dispersion relation becomes
\begin{equation}
\hspace{-3cm} \omega^2 \left(1+\frac{k^2c^2}{\omega_{pe}^2}\frac{1}{\left(1+\frac{m_e}{m_i}\right)}\right)+ i\omega \eta\frac{k^2}{\mu_0}\left(1 - \frac{k_\parallel^2c_{sm}^2}{\omega^2}\right)= k_{\parallel}^2V_A^2\left[\frac{k^2}{k_\perp^2} + k^2\rho_{sm}^2+\frac{k_\parallel^2c_{sm}^2}{\omega^2}\left(\frac{\omega^2}{k_\parallel^2 V_A^2} -\frac{k^2}{k_\perp^2}\right)\right],
\label{eq:emag_femionmom}
\end{equation}
where we note we could rewrite the skin depth contribution as $k^2 c^2/\omega_{pe}^2 = k^2\rho_s^2 V_A^2/v_{te}^2$.
This equation is well behaved at large perpendicular length scales $k_\perp \rightarrow 0$, and we recover the decoupled shear Alfv\'en and ion acoustic waves, as noted in~\cite{scott1997}: $\left(\omega^2 -k_\parallel^2V_A^2k^2/k_\perp^2\right)\left(\omega^2-k_\parallel^2 c_s^2\right)=0$.
If the common approximation to Amp\`ere's law in the strongly magnetised limit $\mu_0 j_\parallel \approx - \nabla_\perp^2 A_\parallel$ is used, $k \rightarrow k_\perp$ in equations~\ref{eq:emag_zem} and~\ref{eq:emag_femionmom}.
Equation~\ref{eq:emag_femionmom} is also now well behaved when $k_\perp$ becomes large.
We can first see this by considering the cold plasma limit, neglecting resistivity, and obtain the electromagnetic modification of equation~\ref{eq:static_esgk}~\cite{lee1987}
\[
\frac{\omega^2k^2m_e}{\Omega_i^2k_\parallel^2m_i} + \left(1+\frac{m_e}{m_i}\right)\left(\frac{\omega^2}{k_\parallel^2V_A^2}-\frac{k^2}{k_\perp^2}\right)=0,
\]
which leads to the inertial Alfv\'{e}n wave dispersion relation,
\begin{equation}
\omega^2 = k_\parallel^2V_A^2/\left[\frac{k_\perp^2}{k^2}+\frac{k_\perp^2c^2}{\omega_{pe}^2}\frac{1}{\left(1+\frac{m_e}{m_i}\right)}\right].
\end{equation}
(It can be of interest to note that in the standard notation of Stix~\cite{stixbook} this result corresponds to the full solution $Sk_\perp^2c^2 + Pk_\parallel^2c^2 -PS\omega^2 =0$,
unlike the quasi-electrostatic mode discussed there which neglects the $PS$ coupling term.)

Now neglecting only the acoustic wave corrections in equation~\ref{eq:emag_femionmom} for simplicity, we find that the modified frequencies of the resistive wave solution~\ref{eq:static_femsoln} are
\begin{eqnarray}
\hspace{-2.5cm}\omega_0^2 &=& k_\parallel^2 V_A^2 \frac{\left(\frac{k^2}{k_\perp^2}+k_\perp^2\rho_s^2\right)}{\left(1+\frac{k^2c^2}{\omega_{pe}^2}\frac{1}{\left(1+\frac{m_e}{m_i}\right)}\right)}\left[1- \frac{\eta^2k^4}{4\mu_0^2k_\parallel^2 V_A^2}\frac{1}{\left(1+\frac{k^2c^2}{\omega_{pe}^2}\frac{1}{\left(1+\frac{m_e}{m_i}\right)}\right)\left(\frac{k^2}{k_\perp^2}+k_\perp^2\rho_s^2\right)} \right], 
\label{eq:emag_femresistivewaveomega} \\
\hspace{-2.5cm}\gamma &=& - \frac{\eta k^2}{2 \mu_0}\frac{1}{\left(1+\frac{k^2c^2}{\omega_{pe}^2}\frac{1}{\left(1+\frac{m_e}{m_i}\right)}\right)},
\label{eq:emag_femresistivewavegamma}
\end{eqnarray}
when the term in the square bracket is positive, for example at low $k_\perp$.
Upon neglecting resistivity, this recovers the standard dispersion relation~\cite{uberoibook} describing the combination of inertial ($T_e \rightarrow 0$) and kinetic Alfv\'{e}n waves
\[
\omega_0^2 = k_\parallel^2 V_A^2 \left(\frac{k^2}{k_\perp^2}+k_\perp^2\rho_s^2\right)/\left(1+\frac{k^2c^2}{\omega_{pe}^2}\frac{1}{\left(1+\frac{m_e}{m_i}\right)}\right),
\]
which reduces in the large $k_\perp$ limit to a wave with electron thermal speed, $\omega_0^2\rightarrow k_{||}^2 v_{te}^2$.
Returning to equation~\ref{eq:emag_femionmom} to consider the limit of finite resistivity and high $k_\perp$, we recover the well-behaved form as in equation~\ref{eq:static_fem},
\begin{equation}
\omega^2 + i \omega \eta \frac{v_{te}^2}{\mu_0 V_A^2 \rho_s^2} = k_\parallel^2v_{Te}^2,
\end{equation}
which has the approximate strongly damped solution at large collisionality $\omega \approx -i(0.51/\tau_{ei}) + (k_\parallel^2v_{te}^2\tau_{ei}/0.51)$.

Note that we do not need to include relativistic corrections
to the electron mass in order to limit the wave speed, except in the cases where $v_{te}$ or $V_A$ exceed $c$. The former occurs at $T_e > 500$keV so would not be relevant to plasma edge simulations, but the latter occurs at $n_0 < 0.66 \times 10^{17}$m$^{-3}$ for a magnetic field of 5T, so may become an issue when simulating the edge of large fusion devices.

\subsubsection{Space charge effects and displacement current}
\label{sec:displacement}

It is reasonable to suppose that if a wave is found which exceeds the speed of light, then we should re-examine the approximations made to 
Maxwell's equations in deriving the reduced MHD equations. In this section we therefore consider the effect of including space charges (breakdown of quasi-neutrality), and then displacement current.

The vorticity equation arises from current continuity (equation~\ref{eq:div_j})
under the quasi-neutrality assumption, that is the charge density $\rho\simeq 0$. This assumption may
break down at small scales and low densities, and charge continuity becomes
\begin{equation}
\nabla\cdot\mathbf{J} = -\frac{\partial \rho}{\partial t} = \epsilon_0\nabla^2\frac{\partial \phi}{\partial t},
\end{equation}
where the Coulomb gauge $\nabla\cdot\mathbf{A}=0$ is used here. This modifies the vorticity equation (\ref{eq:vorticity}) to:
\begin{equation}
\frac{\partial U}{\partial t} = \nabla\cdot\left(\mathbf{b}J_{||}\right), \qquad U \simeq \frac{m_i n_0}{B^2}\nabla_\perp^2\phi + \epsilon_0\nabla^2\phi,
\end{equation}
where the vorticity $U$ has been defined to include the space charge term and can be written as:
\begin{equation}
U = \frac{1}{\mu_0} \left(\frac{1}{V_A^2}\nabla_\perp^2\phi + \frac{1}{c^2}\nabla^2\phi\right).
\end{equation}
This results in the dispersion relation, neglecting finite electron mass and ion acoustic effects:
\begin{equation}
\omega^2 + i\omega\eta\frac{k^2}{\mu_0} = k_{||}^2V_A^2\frac{k^2}{k_\perp^2}\left[ \frac{c^2}{c^2 + V_A^2\left(k^2/k_\perp^2\right)} + k_\perp^2\rho_s^2 \right],
\label{eq:spacedispersion}
\end{equation}
Equation~\ref{eq:spacedispersion} suggests a way to limit the shear Alfv\'en waves to less than light speed at low density,
which will be explored further in section~\ref{sec:conclusion}. It does not however
solve the problem at high $k_\perp$, and waves with arbitrarily high group speed along the magnetic field are still supported.


Since we are not evolving the perpendicular components of the vector potential $\mathbf{A}$, displacement currents in
the perpendicular direction are not included here: they would modify the
perpendicular components of $\mathbf{A}$, but do not lead to motion of charges, and so would not modify charge continuity (vorticity). As we are explicitly evolving only the parallel component of the displacement current, but not 
the parallel component of the magnetic field, the ordinary light wave will not appear in the dispersion relation.
Amp\'ere's law keeping only $A_{||}$ and including the displacement current becomes
\begin{equation}
J_{||} = -\frac{1}{\mu_0}\nabla^2A_{||} + \epsilon_0 \frac{\partial}{\partial t} \partial_{||}\phi + \epsilon_0 \frac{\partial^2 A_{||}}{\partial t^2}.
\end{equation}
Continuing to retain space charge, the vorticity equation becomes
\begin{equation}
  \omega\left[\left(\frac{m_in_0}{B^2} + \epsilon_0\right)k_\perp^2 + \epsilon_0k_{||}^2\right]\phi = k_{||}\left(\frac{k_\perp^2}{\mu_0} - \omega^2\epsilon_0\right)A_{||} + k_{||}^2\epsilon_0\omega\phi.
\label{eq:disp_space}
\end{equation}
Note that the $\epsilon_0 k_{||}^2$ term on the left (from space charge) cancels with the same term on the right (from displacement current).
The dispersion relation then becomes
\begin{equation}
\hspace{-2cm}\omega^2 + i\omega\eta\frac{k^2}{\mu_0}\left(1 -\frac{\omega^2}{k^2c^2}\right) = k_{||}^2V_A^2\frac{k^2}{k_\perp^2}\left[ \frac{c^2}{c^2 + V_A^2\left(k^2/k_\perp^2\right)} + k_\perp^2\rho_s^2\left(1 -\frac{\omega^2}{k^2c^2}\right) \right],
\label{eq:spacejdispdispersion}
\end{equation}
which at large $k_\perp$ without resistivity leads to the same wave as when displacement current is not included (equation~\ref{eq:spacedispersion}). This means that these equations can have waves with speeds
(both group and phase) along the magnetic field greater than the speed of light, without including the electron mass.
However, we see that including the parallel displacement current is not essential to provide effective limitation of the wave speeds in low density regions.



\section{Nonlinear timestep benchmark}
\label{sec:benchmark}

To demonstrate the practical importance of the various dispersion relations derived above, we benchmark nonlinear simulations using different versions of Ohm's law.
We use STORM~\cite{easy2014,rivaetal2019}, which is typical of BOUT++ drift-reduced fluid models, and choose as a test case a simulation of an isolated scrape-off layer (SOL) filament, in simplified slab geometry, based on the cold-ion case in~\cite{ahmedetal2021}.
Isolated filament simulations have a moderate computational cost but include important features of edge/SOL turbulence simulations, being highly non-linear and including sheath boundary conditions.
More details of the model and simulation setup are given in~\ref{app:filament}.

\begin{figure}
    \centering
    \includegraphics{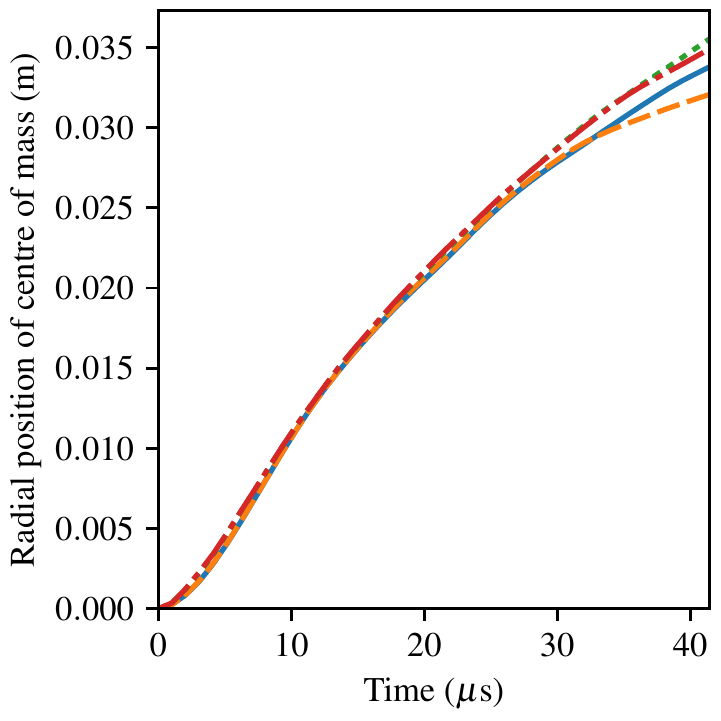}
    \caption{Radial position of the filament centre of mass for electrostatic, zero electron mass (blue solid); electrostatic, finite electron mass (orange dashed); electromagnetic, zero electron mass (green dotted); and electromagnetic, finite electron mass (red dash-dotted) cases.}
    \label{fig:CoM}
\end{figure}

We compare electrostatic and electromagnetic models both neglecting and including electron inertia. The evolution of the filaments simulated with these four different models are essentially identical, due to the very low $\beta\approx 2.6\times 10^{-4}$ in the L-mode SOL conditions of our test case, as shown for example by the radial position of the centre of mass in figure~\ref{fig:CoM}. The slight difference visible between electrostatic and electromagnetic models originates in a small transient at the start of the simulations as the parallel current develops from the initial value of zero, which happens almost instantly in the electrostatic simulations, but at the speed of the Alfv\'en wave in the electromagnetic simulations.

BOUT++ simulations usually use the CVODE implicit time solver from the SUNDIALS suite~\cite{hindmarsh2005}, as the simulations shown in this section do, so do not have a strict Courant-Friedrichs-Lewy (CFL) constraint~\cite{lewy1928} on the timestep, as would be the case if an explicit scheme were used. However, CVODE adapts the timestep used to satisfy relative and absolute tolerance criteria while aiming to minimise the number of iterations, so the resulting timestep is related to the frequency of the fastest mode in the system, as we will now describe.

\begin{table*}

\begin{tabular}{|c|c|c|c|c|}
\hline
Model & $1/|\omega_{\mathrm{analytic}}|$ & timestep & iterations/step & wall-clock time \\
\hline
\hline 
ES, zero-$m_{e}$ & $0.00991\,\mathrm{ns}$ [eq. (\ref{eq:static_zem})] & $0.828\,\mathrm{ns}$ & 8.76 & $30.4\,\mathrm{hrs}$ \\ 
\hline 
ES, finite-$m_{e}$ & $2.47\,\mathrm{ns}$ [eq. (\ref{eq:static_femsoln})] & $0.899\,\mathrm{ns}$ & 3.64 & $11.5\,\mathrm{hrs}$ \\
\hline 
EM zero-$m_{e}$ & $3.18\,\mathrm{ns}$ [eq. (\ref{eq:kaw})] & $7.31\,\mathrm{ns}$ & 6.41 & $3.21\,\mathrm{hrs}$ \\
\hline 
EM, finite-$m_{e}$  & $25.8\,\mathrm{ns}$ [eq. (\ref{eq:emag_femresistivewaveomega})] & $9.25\,\mathrm{ns}$ & 4.35 & $2.36\,\mathrm{hrs}$ \\
\hline 
\end{tabular}

\caption{Timesteps, iteration count and wall-clock time for different electrostatic (ES) and electromagnetic (EM) models.}
\label{tab:timesteps}

\end{table*}

\begin{figure*}
    \centering
    \includegraphics{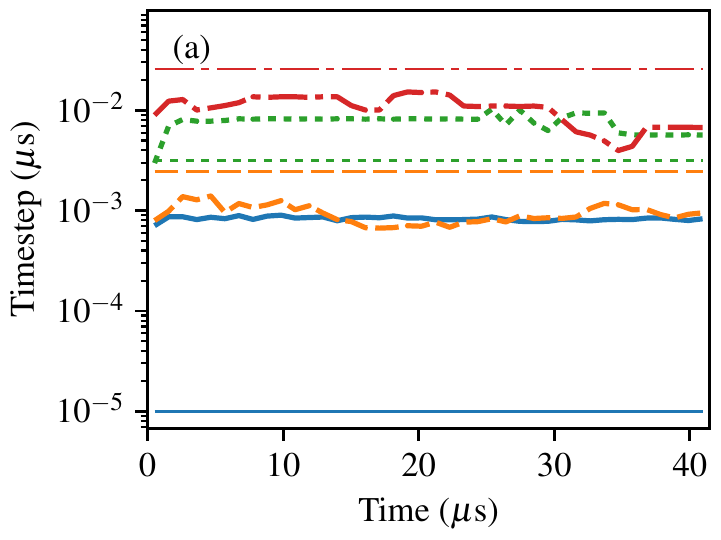}\hfill
    \includegraphics{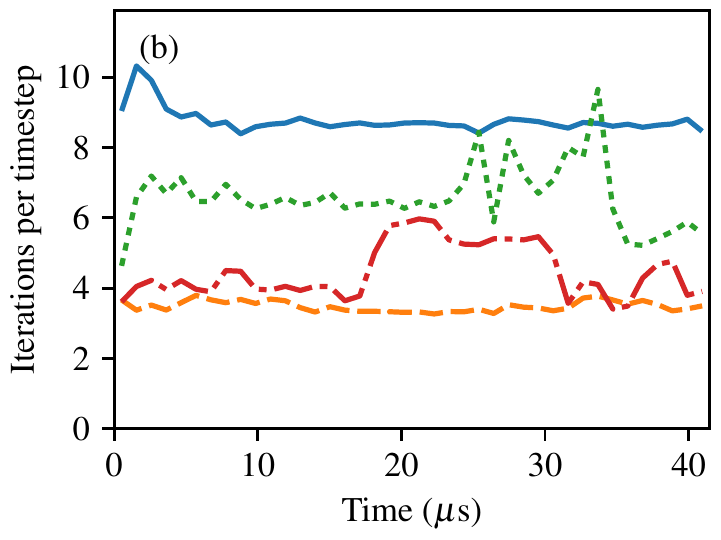}
    \caption{(a) Internal timestep (thick lines) and magnitude of the inverse linear mode frequency (thin, horizontal lines) and (b) number of RHS evaluations per internal timestep for electrostatic, zero electron mass (blue solid, using (\ref{eq:static_zem})); electrostatic, finite electron mass (orange dashed, using (\ref{eq:static_femsoln})); electromagnetic, zero electron mass (green dotted, using (\ref{eq:kaw})); and electromagnetic, finite electron mass (red dash-dotted, using (\ref{eq:emag_femresistivewaveomega})) cases.}
    \label{fig:timestep}
\end{figure*}

We compare the mode frequencies from the dispersion relations presented with the internal timestep, number of iterations per timestep and wall-clock time from the simulations, given in table~\ref{tab:timesteps}. The analytical mode frequencies are evaluated using the background parameters at the midplane of the simulations. The maximum parallel wavenumber present in the simulations $k_{\parallel, \mathrm{max}}=2\pi/2\Delta_\parallel$ is the shortest wavelength that can be represented without aliasing on a grid with parallel spacing $\Delta_\parallel$. Similarly the maximum perpendicular wavenumber $k_{\perp, \mathrm{max}}=2\pi/2\Delta_\perp$ is set by the perpendicular grid spacing $\Delta_\perp$, while the minimum perpendicular wavenumber is set by the box size $k_{\perp,\mathrm{min}}=2\pi/L_\perp$. The maximum damping rate for the resistive, diffusive mode~(\ref{eq:static_zem}) in the electrostatic, zero electron inertia model and the maximum frequency~(\ref{eq:static_femsoln}) for the electrostatic, finite electron inertia model are evaluated with $k_{\perp,\mathrm{min}}$. For the electromagnetic, finite electron inertia model~(\ref{eq:emag_femresistivewaveomega}) also uses $k_{\perp,\mathrm{min}}$, as for the parameters of this test the Alfv\'en speed is higher than the electron thermal speed and we also evaluate it with $k\approx k_\perp$ for consistency with the implemented model. The maximum frequency for the kinetic Alfv\'en wave~(\ref{eq:kaw}) is evaluated with $k_{\perp,\mathrm{max}}$. We see that the average timesteps taken by CVODE in the simulations are ordered
as the inverse mode frequencies, table~\ref{tab:timesteps}, and the timestep and number of iterations per step are fairly consistent in all phases of the simulation, figure~\ref{fig:timestep}. The timesteps for both electrostatic and electromagnetic models with finite electron mass are slightly smaller than the inverse mode frequencies (by a factor between 2 and 3), as would be expected when the simulations are resolving these waves.

The timestep for the strongly damped electrostatic, zero electron mass model is 100 times longer than the inverse of the damping rate~(\ref{eq:static_zem}), showing that the implicit time-solver algorithm used by CVODE is able to step over this mode. The timestep is nearly as long as in the electrostatic, finite electron mass case, although significantly more iterations per timestep are required, leading to a wall-clock time that is three times longer.

The timestep in the electromagnetic, zero electron mass case is slightly longer than the inverse of the maximum mode frequency. Note that this is the maximum $k_\perp$ mode, which will be most affected by numerical dissipation since its wavelength is at the grid scale, and also by collisional dissipation present in the STORM model that was not included in the analytic dispersion relation, so the mode may be expected to have a significant damping rate. It seems that this damping is enough for CVODE to step over the frequency of this mode, at the cost of a higher number of iterations per step than required for the electromagnetic, finite electron inertia model.

The electromagnetic, finite electron inertia model as implemented in STORM requires an additional inversion to solve Amp\`ere's law for the parallel velocities and $A_\parallel$~\cite{ahmedetal2021}, which is not required in the electromagnetic, zero electron inertia model where Ohm's law evolves $A_\parallel$ directly. Such inversions are typically a significant part of the run-time of BOUT++ simulations, here taking $~13\%$ of the run time for the electromagnetic, zero electron inertia model, and $~36\%$ for the electromagnetic, finite electron inertia model. Despite this, the wall-clock time for the electromagnetic, finite electron inertia model is shorter than the zero electron inertia model by a factor of 0.74, due to the longer timestep and lower number of iterations per step.

\section{Conclusion}
\label{sec:conclusion}

We have presented a linear analysis of electrostatic and electromagnetic waves supported by a minimal drift-reduced fluid model. Though simplified,
equations~\ref{eq:density}-\ref{eq:ohms_law} contain the key features
of a wide class of models which are used in the plasma community.
This analysis sheds light on the observation
that once parallel electron
pressure gradients are included in Ohm's law, electrostatic simulations often encounter difficulties.

We recognise the origin of this as
dispersion relations which diverge at small $k_\perp$.
If finite electron mass is not included then this appears as parallel diffusion which becomes faster as $1/k_\perp^2$, whilst if finite electron mass is included
then a parallel wave is found whose speed increases like $1/k_\perp$. 
Use of an electromagnetic Ohm's law without electron mass removes the difficulty at small $k_\perp$, but leads to unphysical behavior due to kinetic Alfv\'{e}n waves at large $k_\perp$, resulting in wave speeds which increase with $k_\perp$ without limit.

A system using an electromagnetic Ohm's law with finite electron mass is found to be well behaved, with parallel wave speeds between the Alfv\'en speed at low $k_\perp$ and the
electron thermal speed at high $k_\perp$ as shown in~\cite{scott1997}. Retaining electromagnetic terms is therefore important for limiting the speed of waves in the system, and we have demonstrated that this can reduce the computational cost of nonlinear simulations by allowing longer timesteps to be taken, even in the very low beta conditions of the L-mode SOL. Elsewhere, it has been found numerically that electromagnetic effects can modify the propagation and stability of plasma blobs~\cite{lee2015} and turbulence~\cite{scott1997} by slowing parallel wave propagation. It has also been previously noted \cite{pfirsch1996} that the electrostatic approximation can lead to inconsistencies and incorrect results even at low $\beta$. 

Plasma edge simulations can include low density regions, resulting in
the Alfv\'en velocity exceeding the speed of light. Adding $\epsilon_0$ to the perpendicular Laplacian term in the vorticity limits $V_A$ to be slower than the speed of light, and can be easily added to existing code. This 
gives the modified equations:
\begin{eqnarray}
\frac{\partial n}{\partial t} &=& \nabla\cdot\left(\mathbf{b}\frac{J_{||}}{e}\right), \label{eq:density2} \\
\nabla\cdot\left(\frac{m_i n}{B^2}\frac{d\nabla_\perp\phi}{dt} + \epsilon_0 \frac{\partial}{\partial t}\nabla_\perp \phi\right) &=& \nabla\cdot\left(\mathbf{b}J_{||}\right), \label{eq:vorticity2} \\
\frac{m_e}{e}\frac{\partial v_{||e}}{\partial t} - \frac{\partial A_{||}}{\partial t}&=&  - \frac{1}{en}\partial_{||}P_e + \partial_{||}\phi  + \eta J_{||}, \label{eq:ohms_law2} \\
&& J_{||} = -\frac{1}{\mu_0}\nabla_\perp^2A_{||},
\end{eqnarray}
which have the dispersion relation:
\begin{equation}
  \omega^2 \left(1 + \rho_s^2k_\perp^2\frac{V_A^2}{V_{te}^2}\right) + i\omega\frac{k_\perp^2\eta}{\mu_0} = k_{||}^2V_A^2\left[\frac{c^2}{c^2 + V_A^2} + k_\perp^2\rho_s^2\right].
\end{equation}
This is now well behaved at high and low $k_\perp$, and in low density regions. Adding the $\epsilon_0\nabla_\perp\phi$ term to the vorticity will have the effect of including the perpendicular electric field energy $\epsilon_0 E_\perp^2$ into the conserved energy of the system, but does not introduce a new transfer channel. This will have the effect of bounding the total energy in perpendicular electric fields, and so may improve numerical stability. In addition we find that the parallel part of $\epsilon_0\nabla^2$ does not need to be included in equation~\ref{eq:vorticity2}, and does not provide a way to introduce parallel coupling into the vorticity equation. This is because it cancels with the displacement current in the divergence of parallel current as discussed in section~\ref{sec:displacement}.

\section*{Acknowledgements}

The author would like to thank J.Madsen, N.R.Walkden and F.I.Parra for many helpful discussions. This work has been carried out within the framework of the EUROfusion Consortium and has received funding from the Euratom research and training programme 2014-2018 under grant agreement No 633053 and the RCUK Energy Programme [grant number EP/T012250/1]. The views and opinions expressed herein do not necessarily reflect those of the European Commission.

\appendix

\section{Filament simulations}
\label{app:filament}

In this appendix we briefly describe the STORM model and simulation setup used for the nonlinear seeded filament simulations discussed in section~\ref{sec:benchmark}. For more details see the cold ion model in~\cite{ahmedetal2021}, and the setup for the simulation shown in figure~1a there, which is identical to the electromagnetic, finite electron mass case here.

\subsection{Models}

The STORM model was originally electrostatic, with finite electron mass~\cite{easy2014}. An electromagnetic, finite electron mass variant was described in~\cite{hoareetal2019}. Our test case is based on the cold ion reference case from~\cite{ahmedetal2021}, which introduced hot ion effects to the electromagnetic, finite electron mass model. For this benchmark we have implemented zero electron mass versions of both electrostatic and electromagnetic models. For all variants the STORM code solves a continuity equation
\begin{eqnarray}
    \frac{\partial n}{\partial t} &=&  -\nabla\cdot\left(\mathbf{b}\;\!n\;\!v_{e\|}\right) - \frac{1}{B}\mathbf{b}\times\nabla\phi\cdot\nabla n \nonumber \\
    && + \nabla\times\left(\frac{\mathbf{b}}{B}\right)\cdot\nabla p_e - n\nabla\times\left(\frac{\mathbf{b}}{B}\right)\cdot\nabla\phi
+ S_n \nonumber \\
    && + \nabla\cdot\left(D_\perp\nabla_\perp n\right),
\end{eqnarray}
where $S_n$ is the density source; an electron temperature equation
\begin{eqnarray}
    \frac{\partial T_e}{\partial t} &= & -v_{e\|}\partial_{\|}T_e - \frac{1}{B}{\mathbf{b}}\times\nabla\phi\cdot\nabla T_e - \frac{2}{3n}\nabla\cdot\left(\mathbf{b}\;\!q_{e\|}\right) \nonumber \\
    && + \frac{2T_e}{3n}\nabla\times\left(\frac{\mathbf{b}}{B}\right)\cdot\left( \nabla p_e - n\nabla\phi + \frac{5}{2}n\nabla T_e \right) \nonumber \\
    && - \frac{2T_e}{3}\nabla\cdot\left(\mathbf{b}\;\!v_{e\|}\right) + \frac{2}{3}\left(v_{i\|} - v_{e\|}\right) \left(\eta J_\parallel - \frac{0.71}{e}\partial_\parallel T\right) \nonumber \\
    && + \frac{2S_E}{3n} + \frac{v_{e\|}^2S_n}{3em_e n} - \frac{T_e S_n}{n} + \frac{2}{3n}\nabla\cdot\left(\kappa_{e\perp}\nabla_\perp T_e\right),
\end{eqnarray}
where the parallel electron thermal conduction is $q_{e\parallel}=-3.16en T \tau_{ei} \partial_\parallel T/m_e - 0.71n T\left(v_{i\|} - v_{e\|}\right)$ and $S_E$ is the energy source; and a vorticity equation
\begin{eqnarray}
    \frac{\partial \Omega}{\partial t} &= & -\nabla\cdot\left( \frac{1}{B}\mathbf{b}\times\nabla\phi\cdot\nabla\boldsymbol{\omega} \right) - \nabla\cdot\left( \partial_{\|}\left(v_{i\|}\boldsymbol{\omega}\right) \right) \nonumber \\
    && + \nabla\cdot\left(\mathbf{b}\;\!J_\|\right) + e\nabla\times\left(\frac{\mathbf{b}}{B}\right)\cdot\nabla p_e + \nabla\cdot\left(\mu_\Omega\nabla_\perp\Omega\right),
\end{eqnarray}
where for this non-linear model we use a generalised vorticity without Boussinesq approximation $\Omega=\nabla\cdot\boldsymbol{\omega}$, $\boldsymbol{\omega}=en\nabla_\perp\phi / \Omega_i B$. The perpendicular dissipation parameters $D_\perp$, $\kappa_\perp$ and $\mu_\Omega$ take small, classical values as described in~\cite{ahmedetal2021} and do not have a significant influence on the results, being retained mainly to ensure numerical stability.

The differences between variants are in the parallel momentum equations. We use the parameters $\alpha$ and $\mu$, where
\begin{eqnarray*}
    \alpha &=& \cases{
        0 & for electrostatic cases, \\
        1 & for electromagnetic cases,
    } \\
    \mu &=& \cases{
        0 & for zero electron mass, \\
        1 & for finite electron mass,
    }
\end{eqnarray*}
to express the ion parallel momentum equation as
\begin{eqnarray}
    \frac{\partial}{\partial t}\left(v_{i\|} + \mu\alpha\frac{e}{m_i}A_\|\right) &= & -v_{i\|}\partial_{\|}v_{i\|} - \frac{1}{B}\mathbf{b}\times\nabla\phi\cdot\nabla v_{i\|} \nonumber \\
    && - \frac{e}{m_i}\partial_{\|}\phi - \frac{e\eta}{m_i}J_\parallel + \frac{0.71}{m_i}\partial_\parallel T - \frac{v_{i\|} S_n}{n}
\end{eqnarray}
and Ohm's law as
\begin{eqnarray}
    \mu\frac{\partial}{\partial t}\left(v_{e\|} - \alpha\frac{e}{m_e}A_\|\right) &= & -\mu\left(v_{e\|}\partial_{\|}v_{e\|} + \frac{1}{B}\mathbf{b}\times\nabla\phi\cdot\nabla v_{e\|} - \frac{v_{e\|}S_n}{n}\right) \nonumber \\
    && - \frac{e}{m_e n}\partial_{\|}p_e + \frac{e}{m_e}\partial_{\|}\phi + \frac{e\eta}{m_e}J_\parallel - \frac{0.71}{m_e}\partial_\parallel T. \nonumber \\
\end{eqnarray}

\subsection{Isolated filament simulation setup}

The simulations are performed in slab geometry with a constant magnetic field of $B=0.5\,\mathrm{T}$.
The effect of non-uniform magnetic field is retained only through the curvature terms $\nabla\times\left({\bf b}/B\right)\cdot\nabla = 2\left(B R_c\right)^{-1}\nabla_z$ with radius of curvature $R_c=1.5\,\mathrm{m}$.
The spatial grid has $240\times64\times128$ points in the radial, parallel and binormal directions $x$, $y$ and $z$. The grid size is $L_x\times L_\parallel\times L_z = 93.75\rho_{s0}\times11000\rho_{s0}\times50\rho_{s0}$, where $\rho_{s0}=\rho_s(T=20\,\mathrm{eV})=1.29\,\mathrm{mm}$, giving a perpendicular grid spacing of $0.391\rho_{s0}$ and a parallel grid spacing of $172\rho_{s0}$. The parallel boundaries use Bohm sheath boundary conditions: $v_{i\parallel,sh} = \pm\sqrt{eT_{sh}/\left(m_i + m_e\right)}$, $v_{e\parallel,sh} = \pm\sqrt{m_i eT_{sh} / 2\pi m_e\left(m_i + m_e\right)}\exp\!\left(-\phi_{sh}/T_{sh}\right)$ and $q_{e\parallel,sh} = \left(2 - 0.5\ln\!\left(2\pi m_e/m_i\right)\right)n_{sh}T_{sh}v_{e\parallel,sh} - 5p_{e,sh} v_{e\parallel,sh}/2 - m_e n_{sh} v_{e\parallel,sh}^3/2e$. The perpendicular boundaries are far enough from the filament to have little effect: Neumann boundary conditions are used in the radial direction for all variables except $\phi$, which is set equal to its background value; periodic boundary conditions are used in the binormal direction.

A steady background plasma is created by balancing the sinks at the sheaths with sources $S_n = S_{n0}\left[\exp\!\left(20\left(y/L_\parallel-1/2\right)\right) + \exp\!\left(-20\left(y/L_\parallel+1/2\right)\right)\right]$ and $S_E = S_{E0}\exp\!\left(-10\left|y\right|/L_\parallel\right)$, whose prefactors are adjusted so that the density and temperature at the mid-point of the simulation domain are $8\times 10^{18}\,\mathrm{m^{-3}}$ and $20\,\mathrm{eV}$.

The filament simulations are run by adding a density perturbation on top of the background
\begin{eqnarray}
    \Delta_n &=& A \frac{1}{2} \left[1 - \tanh\left(({\textstyle\frac{y - L_\parallel/4}{L_\parallel/8}}\right)\right] \frac{1}{2} \left[1 - \tanh\left({\textstyle\frac{-y - L_\parallel/4}{L_\parallel/8}}\right)\right] \exp\!\left({\textstyle -\frac{x^2 + z^2}{\delta_\perp^2}}\right), \nonumber \\
\end{eqnarray}
with amplitude $A=1.6\times 10^{19}\,\mathrm{m^{-3}}$ and perpendicular width $\delta_\perp = 5\rho_{s0}$, and allowing the simulation to evolve for $1000 \,\Omega_i^{-1} = 41.5\,\mathrm{\mu s}$.

\section*{References}
\bibliography{references}
\bibliographystyle{unsrt}

\end{document}